\documentclass[epj]{svjour}

\usepackage{amsmath}
\usepackage{amsfonts}
\usepackage{amssymb}

\usepackage{units}
\usepackage{psfrag}

\usepackage{graphicx}

\newcommand{\norm}{\mathcal{N}}
\newcommand{\energy}{E}
\newcommand{\new}[1]{#1}

\begin{document}

\title{Re-localization due to finite response times in a nonlinear Anderson chain}

\author{Mario Mulansky\inst{1} \and Arkady Pikovsky\inst{1}}
\institute{Department of Physics and Astronomy, Potsdam University, 14476 Potsdam, Germany, EU
}

\abstract{
We study a disordered nonlinear Schr\"odinger equation with an additional relaxation process having a finite response time $\tau$.
Without the relaxation term, $\tau=0$, this model has been widely studied in the past and numerical simulations showed subdiffusive spreading of initially localized excitations.
However, recently Caetano et al.\ (EPJ.~B~\textbf{80}, 2011) found that by introducing a response time $\tau > 0$, spreading is suppressed and any initially localized excitation will remain localized.
Here, we explain the lack of subdiffusive spreading for $\tau>0$ by numerically analyzing the energy evolution.
We find that in the presence of a relaxation process the energy drifts towards the band edge, which enforces the population of fewer and fewer localized modes and hence leads to re-localization.
The explanation presented here relies on former findings by the authors et al.\ (PRE \textbf{80}, 2009) on the energy dependence of thermalized states.
}


\maketitle
\enlargethispage{\baselineskip}

Localization in linear, disordered systems is a well established phenomenon also known as Anderson localization~\cite{anderson:1958}.
One interesting aspect is the interplay between localization due to disorder, and nonlinearity, which by intuition should lead to thermalization in high dimensional systems.
This is mostly formulated as a spreading problem of initially localized states and has first been addressed mainly numerically~\cite{Pikovsky-Shepelyansky:08}, but lately also experimental studies~\cite{Nature.446.52,Nature.6.354} as well as mathematical treatments~\cite{Basko-10} were established.
Numerical results showed a sub-diffusive spreading of an initially localized wave packet over a large timescale up to $t\sim 10^9$~\cite{flach:024101,Mulansky-Pikovsky-10}.
Lately, the spreading behavior was also related to properties of high-dimensional chaos~\cite{mulansky:2011}.

\begin{figure*}[t!]
 \centering
 \psfrag{tau = 0.1}{}
 \psfrag{tau = 0.001}{}
 \includegraphics[width=0.48\textwidth]{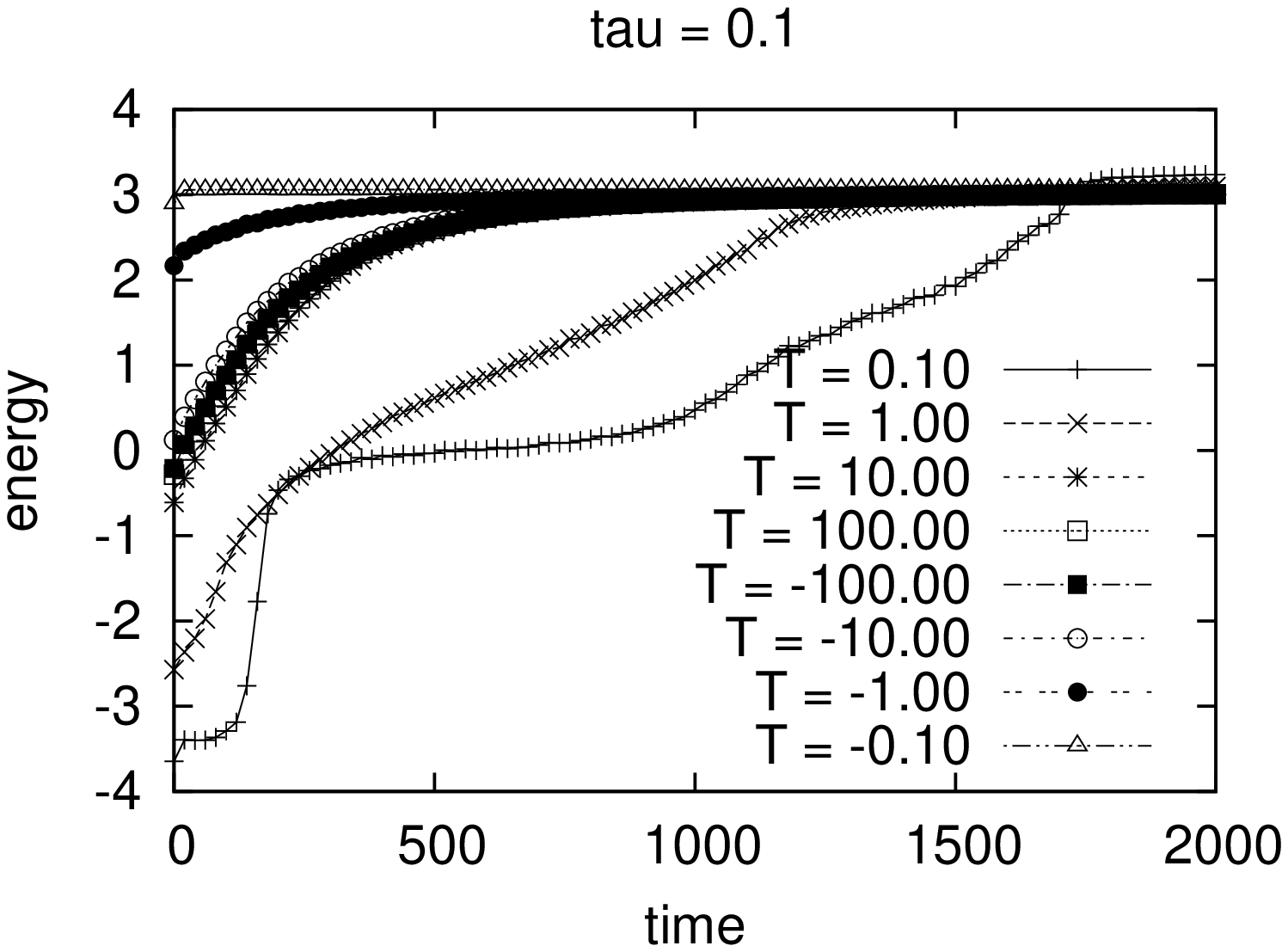} \hfill
 \includegraphics[width=0.48\textwidth]{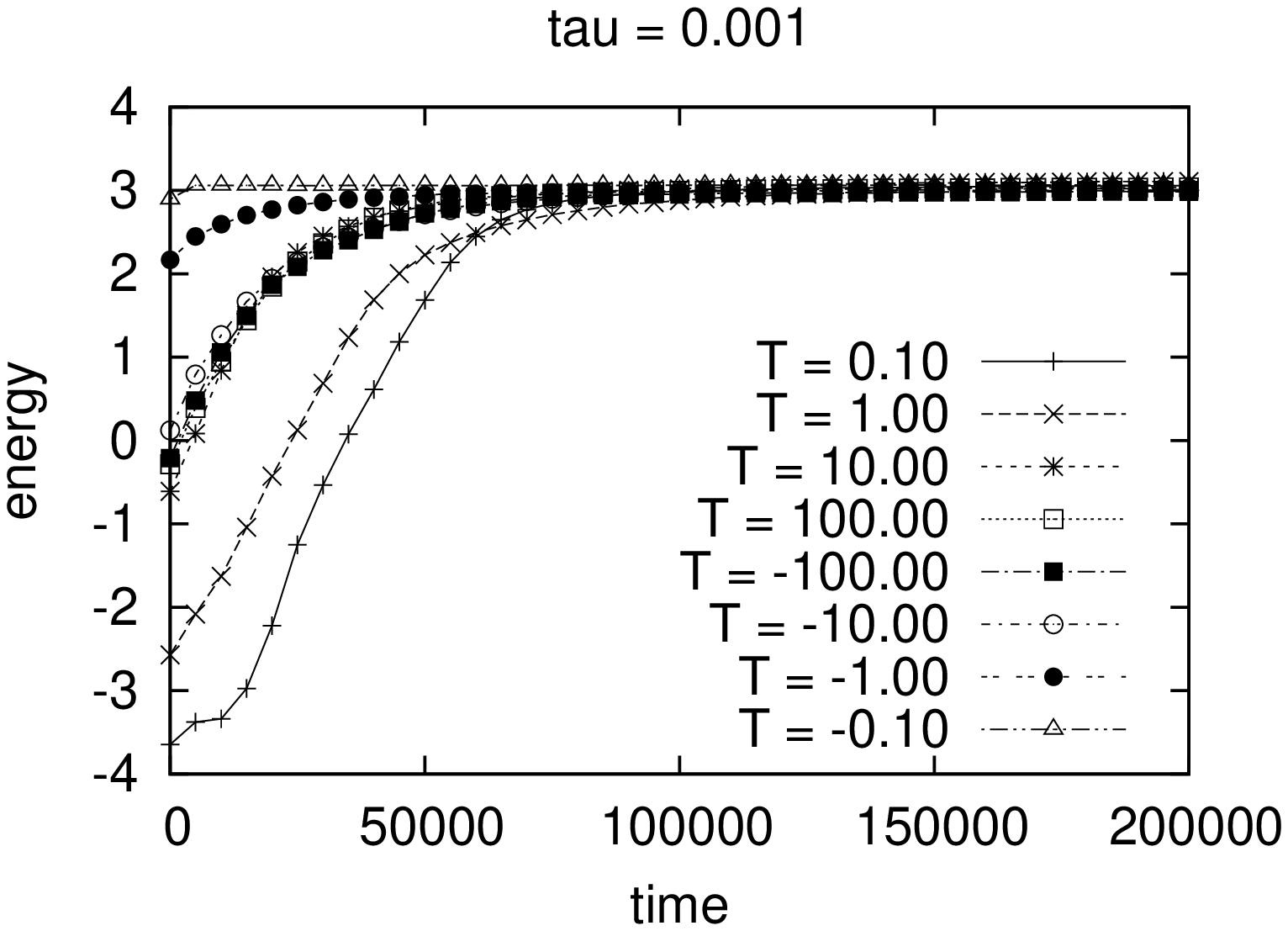} 
 \caption{Energy of the state as function of time for $\beta=1.0$ with $\tau = 0.1$ (left) and $\tau = 0.001$ (right). Shown is the time evolution of an initially thermalized state (maximal entropy, Gibbs distribution, see text) in a lattice with $N=32$ sites for a given temperature $T = -100 \dots 100$.
}
 \label{fig:en_time}
\end{figure*}

There are a number of models where disorder and nonlinearity can be studied, for example Hamiltonian oscillator chains~\cite{mulansky:2011_2}, the Klein-Gordon model~\cite{flach:024101} or coupled map lattices~\cite{shepelyansky:1993}.
However, here we use the Discrete Anderson Nonlinear Schr\"odinger Equation (DANSE) and we start with a short summary of its properties found in previous studies~\cite{mulansky:2009}.
The equations of motion for this model are:
\begin{equation} \label{eqn:danse_eq_motion}
  i\dot\psi_n = V_n \psi_n + \psi_{n-1} + \psi_{n+1} + \beta |\psi_n|^2 \psi_n
\end{equation}
and can be derived from the Hamiltonian:
\begin{equation} \label{eqn:danse_hamiltonian}
 H = \sum_n V_n|\psi_n|^2 + \psi_n\psi_{n+1}^* + \psi_n^*\psi_{n+1} + \frac\beta2 |\psi_n|^4.
\end{equation} 
$\psi_n$ is the complex valued field amplitude, $V_n$ is a random potential choosen iid.\ from $(-W/2\dots W/2)$ with typically $W = 4.0$ and $\beta$ describes the nonlinear strength with a typical value of $\beta=1$.
We emphasize that this model exhibits two conserved quantities: norm ${\norm = \sum |\psi_n|^2 = 1}$ and energy $\energy = H$.
While the norm can be scaled to ${N=1}$ by a rescaling of $\beta$, the energy is an important quantity governing the behavior of the system.
This energy dependence has been investigated in~\cite{mulansky:2009} and we shortly summarize the results, as they will help understanding the modified model studied below.
In the linear case, $\beta = 0$, the Hamiltonian~\eqref{eqn:danse_hamiltonian} can be diagonalized: ${H = \sum_k \epsilon_k c_k^* c_k}$ whith energy eigenvalues $\epsilon_k$ and eigenfunctions $\varphi_k$ where $\psi_n = \sum_k c_k \varphi_{k,n}$.
Due to the disorder $V_n$, $\epsilon_k$ are randomly distributed in the interval $[-W/2-2 , W/2+2]$ and the $\varphi_k$ are exponentially localized in space~\cite{anderson:1958} (for a review see e.g.~\cite{RevModPhys.57.287}).
Adding nonlinearity $\beta \neq 0$ one still can write the system in terms of the linear eigenfunctions, but now an additional 4-mode coupling term  
will appear in the diagonalized Hamiltonian allowing for energy/norm exchange between linear modes, hence spreading.

\enlargethispage{2\baselineskip}

In~\cite{mulansky:2009} it was shown that the energy is a crucial parameter determining the efficiency of spreading.
If one initially excites an eigenmode with energy at the band edge, no spreading can appear because of energy and norm conservation.
There simply does not exist a state with such a norm and energy value other than the localized eigenmode at the band edge.
If, on the other hand, time evolution is started from an excitation with an energy close to the band center $E\approx0$, the conservation laws do not prevent the system from reaching extended states and typically spreading is observed numerically.
This can be formulated more rigorously by considering thermalization in such a system in terms of a Gibbs distribution~\cite{mulansky:2009}.
It should be noted that a thermalized state is usually also maximally spread as a maximum Gibbs-entropy corresponds to a maximum number of excited lattices sites or eigendmodes, respectively.
Denoting $c_k$ the amplitude of the $k$-th eigenmode such that $\psi_n = \sum_k c_k \varphi_{k,n}$ we find for the thermalized state:
\begin{equation} \label{eqn:gibbs}
  |c_k|^2 = \mathrm{e}^{-\epsilon_k / T }/Z , \qquad \text{with}\qquad Z = \sum_k \mathrm{e}^{-\epsilon_k / T },
\end{equation}
where $T\in(-\infty,\infty)$ is the dimensionless effective temperature.
Note that for a specific disorder realization, fixing $T$ is equivalent to fixing the energy $E$, so that temperature values $\pm\infty$ correspond to energies at the band center while $T=\pm 0$ belong to energies at the band edges.
From~\eqref{eqn:gibbs} it is seen that in the limit $T\to\pm\infty$ the thermalized state is an uniform excitation of all eigenmodes, hence extended, while for $T \to \pm 0$ only the eigenmode with largest/smallest eigenvalue gets excited which results in a localized final state.
So changing the energy from the band center towards the band edge makes the thermalized state becoming more and more localized.
This has been found in~\cite{mulansky:2009} and will now be used to explain the results in a modified model.

Recently, Caetano et al.\ introduced a modification of the DANSE model that consists of an additional relaxation process introduced on top of the nonlinearity~\cite{caetano:2011}:
\begin{equation} \label{eqn:model}
\begin{aligned} 
  i\dot\psi_n &= V_n \psi_n + \psi_{n-1} + \psi_{n+1} + \beta f_n \psi_n \\
  \tau\dot f_n &= |\psi_n|^2 - f_n.
\end{aligned}
\end{equation}
It was found numerically by Caetano et al.\ that in this system, any non-zero relaxation time $\tau>0$ surpresses spreading.
However, up to now no explanation for this absence of spreading has been given.
We now are going to present such an explanation.

We first note that by introducing the relaxation process we destroy the energy conservation valid for the original DANSE model ($\tau=0$ , $f_n(t) = |\psi_n(t)|^2$).
However, the norm $\norm = \sum |\psi_n|^2$ is still conserved here and we again fix $\norm = 1$.
The energy becomes time dependent and calculates as:
\begin{equation}
 E(t) = \sum_n V_n|\psi_n|^2 + 2\Re(\psi_n\psi_{n+1}^*) + \frac\beta2 f_n(t) |\psi_n|^2.
\end{equation} 

\enlargethispage{\baselineskip}

Numerical simulations presented below show that for any $\tau>0$ the energy increases with time if $\beta > 0$, while it decreases for $\beta < 0$.\footnote{Note that we used $f_n(0)=0$ in all simulations, for other initial values $f_n(0)$ the energy behavior might also depend on the sign of the term $|\psi_n|^2-f_n$.}
But as $\norm$ is still conserved in this system, the energy can not increase (decrease) unbounded.
Norm conservation still enforces a strict bound on the energy $E$, namely $-2-W/2+\beta/2 < E < 2+W/2+\beta/2$.

In a first numerical study we fixed a lattice with ${N=32}$ sites and used a thermalized state as given by \eqref{eqn:gibbs} with a given temperature $T$ as an initial condition \new{for $\psi_n$ and $f_n(0)=0$}.
Starting from such an initial condition  for a given temperature and fixing $\beta=+1$, we integrated the equations of motion \eqref{eqn:model} using the Dopri5-Runge-Kutta method of order 5 with error control (tolerance $10^{-8}$) and dense output \cite{odeint}.
This is not the most efficient numerical method for a stiff system like our modfied DANSE, but for the results presented here the computational effort did not require us to apply more sophisticated routines.
Exemplary results for $\tau=0.1$ and $\tau=0.001$ are shown in Fig.~\ref{fig:en_time}.
For all tested values of $\tau=0.001\dots 100$ we found an energy growth during the time evolution up to a saturation value $E_s \approx 3.0$, which corresponds to the maximal energy eigenvalue, hence the upper band edge.
Obviously, the time-scale at which this energy drift takes place is governed by $\tau$.
As seen in Fig.~\ref{fig:en_time}, changing $\tau$ by two orders of magnitude also changes the time-scale of the energy drift by roughly two orders of magnitude.
Numerical results for $\beta=-1$ are very similar but now the energy decreases down to $E_s \approx -3$ for all values of $\tau$ and all tested initial conditions.
\begin{figure*}[t]
 \psfrag{label1}[r][r]{$\beta = +1$}
 \psfrag{label2}[r][r]{$\beta = -1$}
 \centering
 \includegraphics[width=0.48\textwidth]{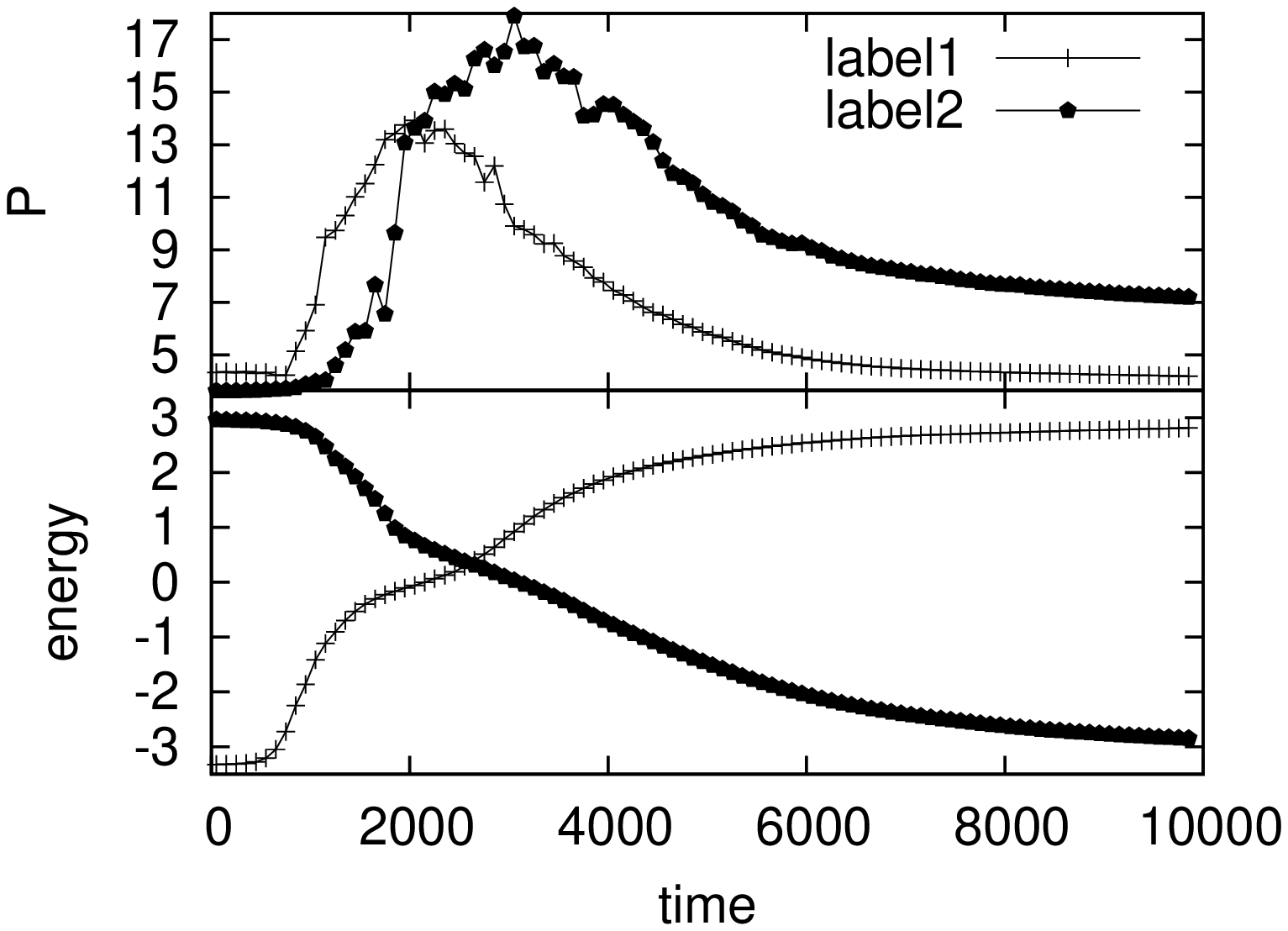} \hfill
 \includegraphics[width=0.48\textwidth]{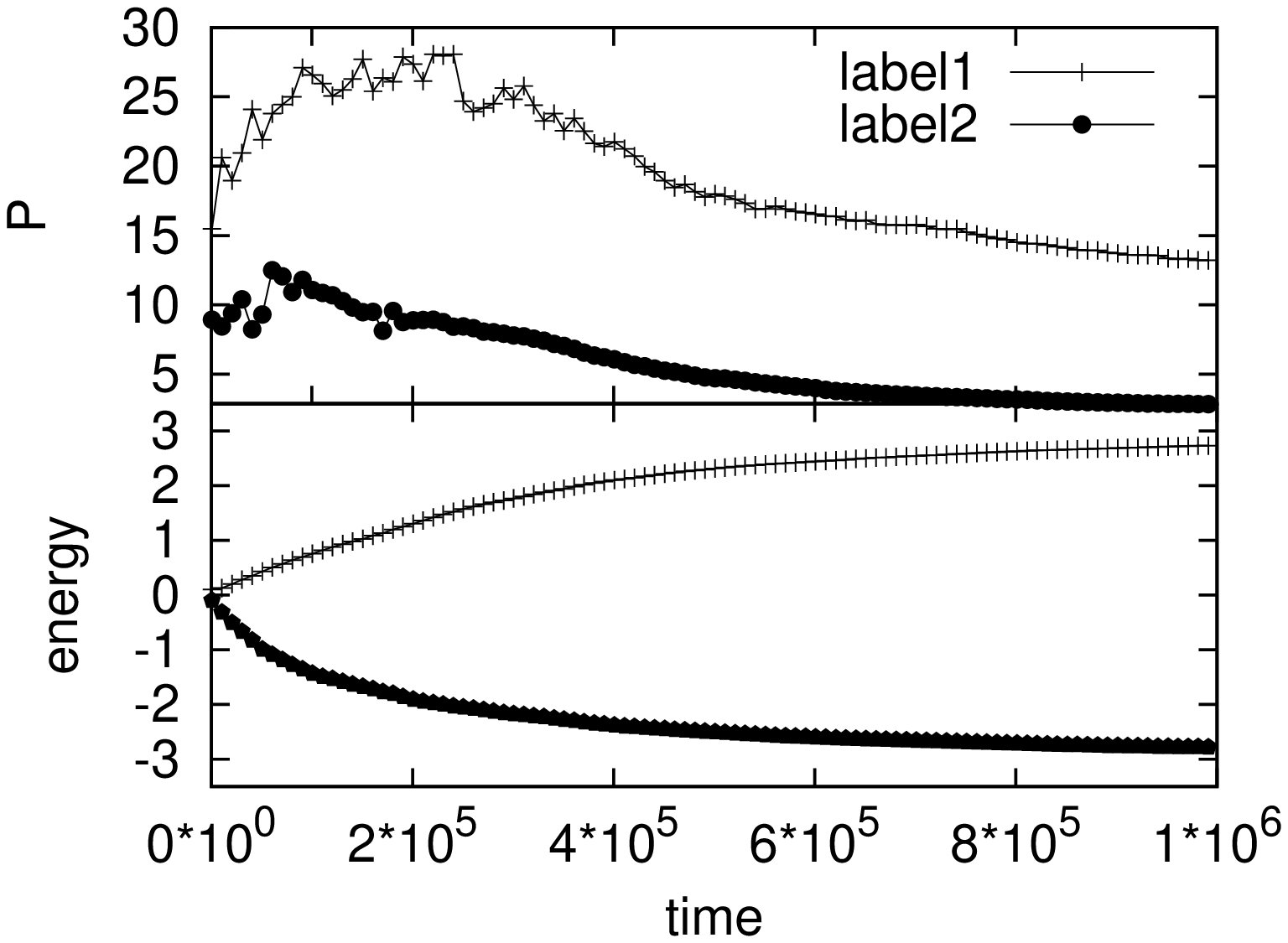} \\ 
 \caption{Spreading of initial eigenstates with lowest ($\beta=1$) and highest ($\beta=-1$) energy value for $\tau=0.01$ (left panel). In the right panel we initialized the system with eigenstates from the band center ($E\approx 0$) and started time evolution with $\beta=\pm1$ using $\tau=0.0001$. Shown are participation number $P=1/\sum|\psi_n|^4$ and energy as functions of time.}
 \label{fig:spread}
\end{figure*}

\enlargethispage{\baselineskip}

From the consideration above we can now conclude that even if thermalization occurs, the final state must be localized as it will always have an energy close to the band edge.
So the numerical observation of the drift of the energy together with the previously found energy dependence of thermalized states~\cite{mulansky:2009} explains the lack of spreading in systems with finite response time.
\new{We note that the localization of high (low) energy modes can also be understood in terms of a self-trapping mechanism.
Indeed, if the energy is shifted out of the spectrum of energy eigenvalues of the linear problem ($\beta=0$), then by adding nonlinearity a localized mode is created and spreading is prevented~\cite{kopidakis:2008}.}

\enlargethispage{\baselineskip}

In a second numerical experiment we revisit the spreading problem for system~\eqref{eqn:model} to further visualize the interplay between spreading and energy drift.
We took a lattice of $N=256$ sites and choose either the eigenmode with lowest/highest energy or an eigenmode from the band center as initial condition \new{for $\psi_n$ and again $f_n(0)=0$}.
Then we ran the time evolution with $\beta=\pm1.0$ and measured the spreading as well as the energy $E$.
As measure of spreading we chose the participation number $P=1/\sum |\psi_n|^4$ that is a widely used quantity in such problems as it roughly counts the number of excited sites.
Large numbers of $P$ correspond to spreading while small $P$ indicate localization.
The results are shown in Fig.~\ref{fig:spread}.
As is seen there, starting from the highly localized eigenstates at the band edge the relaxation process increases ($\beta = +1$) or decreases ($\beta=-1$) the energy which initially leads to spreading.
When the energy crosses zero the wavefunction is maximally delocalized, but a further increase/decrease leads back to localization until the energy reaches the other edge of the energy band and the process saturates in a localized state.
Hence, the energy drift towards the band edges induces re-localization of the wave function.
The same happens when starting from states with initial energies at the band center $E\approx0$ as also shown in Fig.~\ref{fig:spread}.
There, some initial spreading takes place due to nonlinear interactions, but finally also re-localization appears due to the energy drift.

To summarize, we found an explanation for the surpression of subdiffusive behavior in a modified DANSE model with an additional relaxation process with response time $\tau > 0$.
Our reason is based on the numerical observation of an energy drift induced by the relaxation process.
This energy drift continues up (down) to the energy bound as given by norm conservation in the system.
Using previous results about thermalization properties in the DANSE model stating that states close to the band edges must be localized~\cite{mulansky:2009}, we found the explanation for the effect of re-localization observed here and in a previous study~\cite{caetano:2011}.
We note that our explanation is based on the numerical observation of an energy drift.
A rigorous analytic treatment of the dynamic equations should be done to obtain mathematical results showing the drift of the energy, but this is left for future work.
To our opinion, the importance of the energy on thermalization as described in \cite{mulansky:2009} is crucial for understanding such modified DANSE models.

\enlargethispage{\baselineskip}

\end{document}